\documentclass[conference]{IEEEtran}
\IEEEoverridecommandlockouts
\usepackage{cite}
\usepackage{amsmath, amssymb, amsfonts, subcaption}
\usepackage{algorithmic}
\usepackage{graphicx}
\usepackage{textcomp}
\usepackage{xcolor}

\DeclareMathOperator*{\argmin}{arg\min}
\def\BibTeX{{\rm B\kern-.05em{\sc i\kern-.025em b}\kern-.08em
    T\kern-.1667em\lower.7ex\hbox{E}\kern-.125emX}}
\begin{document}

\title{LoRa-based Over-the-Air Computing for Sat-IoT
\thanks{This work is part of the project IRENE (PID2020-115323RB-C31), funded by MCIN/AEI/10.13039/501100011033 and supported by the Catalan government through the project SGR-Cat 2021-01207.}
}

\author{\IEEEauthorblockN{Marc Martinez-Gost \IEEEauthorrefmark{1}\IEEEauthorrefmark{2}, Ana Pérez-Neira\IEEEauthorrefmark{1}\IEEEauthorrefmark{2}\IEEEauthorrefmark{3}, Miguel Ángel Lagunas\IEEEauthorrefmark{1}\IEEEauthorrefmark{2}}
\IEEEauthorblockA{
\IEEEauthorrefmark{1}Centre Tecnològic de Telecomunicacions de Catalunya, Spain\\
\IEEEauthorrefmark{2}Dept. of Signal Theory and Communications, Universitat Politècnica de Catalunya, Spain\\
\IEEEauthorrefmark{3}ICREA Acadèmia, Spain\\
\{mmartinez, aperez, malagunas\}@cttc.es}
}

\maketitle

\begin{abstract}
Satellite Internet of Things (Sat-IoT) is a novel framework in which satellites integrate sensing, communication and computing capabilities to carry out task-oriented communications. In this paper we propose to use the Long Range (LoRa) modulation for the purpose of estimation in a Sat-IoT scenario. Then we realize that the collisions generated by LoRa can be harnessed in an Over-the-Air Computing (AirComp) framework. Specifically, we propose to use LoRa for Type-based Multiple Access (TBMA), a semantic-aware scheme in which communication resources are assigned to different parameters, not users. Our experimental results show that LoRa-TBMA is suitable as a massive access scheme, provides large gains in terms of mean squared error (MSE) and saves scarce satellite communication resources (i.e., power, latency and bandwidth) with respect to orthogonal multiple access schemes. We also analyze the satellite scenarios that could take advantage of the LoRa-TBMA scheme. In summary, that angular modulations, which are very useful in satellite communications, can also benefit from AirComp.
\end{abstract}

\section{Introduction}
The future 6G (6th Generation) communication is expected to integrate the terrestrial systems and non-terrestrial satellite constellations seamlessly, which is known as a 3D wireless architecture. Low Earth Orbit (LEO) satellite networks will coordinate joint sensing, communication and computing capabilities. The sensing capabilities (or terrestrial IoT devices) allow the satellites to gather data and the computational resources allow to reduce the communication and computational burden on the ground segment. This paradigm is called Satellite Internet of Things (Sat-IoT), which provide the envisioned Intelligence as a Service in the future 6G networks \cite{Qu2017, Akyildiz2019}.

The consideration of joint communication and computing schemes leads to the development of task-oriented communications, this is, communicating to accomplish the goal of a certain task, and not for the sake of transmitting information reliably \cite{Nazer2007}. Semantic communications is an intimately related topic, in which the transmitted information is designed according to the meaning of the information with respect to a given task \cite{Qin2021}.

The goal of the present paper is to use the Long Range (LoRa) modulation \cite{Chiani2019} for the purpose of estimation in Sat-IoT. In our previous work \cite{Gost23} we developed DCT-FM, a modulation for the task of estimation and function approximation through a communication channel. By exploiting the Discrete Cosine Transform (DCT) to design the waveform and approximate the mathematical function, this results in a low cost and high precision scheme in terms of communication and computing. Specifically, we note that LoRa can be derived as a DCT-FM for computing. This modulation is suitable for the satellite environment and complies with the requirements of satellite communications: stringent energy constraints, limited Line of Sight (LoS) transmission, high Doppler shift, etc.

Then we generalize the estimation problem for multiple users. In this respect, the collisions generated by LoRa in the simultaneous transmission of the same information can be leveraged as an scheme that takes advantage of an Over-the-Air computing (AirComp) access scheme \cite{Nazer2007}. Particularly, LoRa is suitable for implementing Type-based Multiple Access (TBMA), a semantic-aware scheme that allows the receiver to estimate a parameter from the concurrent transmission of distributed noisy observations \cite{Mergen2006}. TBMA is suitable for semantic communications, as it assigns resources according to how informative the datum of a transmitter is with respect to all the other data. In this respect, TBMA relies on AirComp and it inherits efficiency in terms of power, latency and bandwidth. Constant envelope modulations, such as LoRa, are very useful in satellite communications. This paper innovates showing that these modulations can also benefit from an AirComp access.
Finally, we present the scenarios under which TBMA may be useful in the satellite environment and provide experiments evincing that the performance of TBMA is substantially better with respect to orthogonal schemes in terms of mean squared error (MSE) and resource consumption.

The remaining part of the paper proceeds as follows: Section II introduces the system model and the LoRa modulation for estimation. Then, section III focuses on multiple users and proposes LoRa-TBMA for estimation. Finally, Section IV complements the theoretical analysis with experimental results and Section V concludes the paper.

\section{Frequency Modulation for Computing}
\label{sec:method}
\subsection{System Model}
Consider a signal $x(t)$, which remains constant over an observation time $T$. For instance, $T$ may be associated to the changing nature of a parameter in a sensing environment or the time it takes to process data in a computation scenario. Thus, throughout the rest of the paper we drop the time index and work with $x$. In a point-to-point setting, we consider that the transmitter acquires $x$ and, using an analog-to-digital converter (ADC), rounds it to the nearest integer: $Q: x \rightarrow m$, this is, $m\in [0, N-1]\subset\mathbb{N}$. Throughout the rest of the paper we refer to $m$ as the measurement.

The goal of the system is to estimate a parameter $\theta$ from the measurement. As in our previous work, we focus on computing a function $\theta=f(m)$. We refer the reader to \cite{Gost23} for a description of the functions that are suitable for this scenario.
The transmitter modulates the datum $m$ and sends it over the channel, which is modelled as a large scale fading $h$. Note that these scenarios (inter-satellite and ground-space links) comprise LoS propagation with free space path loss and other sources of ground-space losses (e.g., atmospheric absorption, rain attenuation, polarization, etc.), but not multipath. The noise $w$ has spectral density equal to $N_o/2\, [W/Hz]$.
At the receiver side, demodulation allows to recover $m$, which is processed to obtain an estimate of the function $\hat{f}(m)$. The performance of the system is measured in terms of the mean MSE between the original function and the reconstruction, as it will be formulated later on.

In this work we consider a low-pass equivalent and discrete-time model. In this respect, the bandwidth of signal $x$ is $W=1/T$ and we choose a sampling frequency of $f_s=N/T$, delivering $N\geq2$ samples per measurement.


\subsection{DCT-FM: A Modulation for Computing}
In \cite{Gost23} we presented the following baseband modulation,
\begin{equation}
z_{ag}[n] = A_c\sqrt{\frac{2}{N}}\sum_{k\in \mathcal{K}} F_k \cos\left(\frac{\pi k(2m+1)}{2N}n\right),
\label{eq:prev_dct_mod}
\end{equation}
for $n=0,\dots, N-1$, where $A_c$ is the amplitude of the carrier, $F_k$ are the DCT coefficients for $k\in\mathcal{K}$. This expression comes from the inverse DCT (iDCT), incorporating a time index $n$. By computing the DCT of $z_{ag}$ in the time domain (i.e., index $n$) the receiver obtains the noisy coefficients $\tilde{F}_k$, from which it can recover both the measurement $m$ and reconstruct the function $\hat{f}(m)$. We refer the reader to our previous work for the details in signal demodulation and how to estimate the function \cite{Gost23}.


In the context of satellite communications this modulation does not reveal good properties that adapt to the communication channel. Mainly because it is not only a frequency, but also an amplitude modulation where the tones decay very fast (around 12 dB per coefficient). Therefore, this modulation is not suited for large attenuation channels as in satellite systems. In this respect we provide the following variation of \eqref{eq:prev_dct_mod}:
\begin{equation}
z[n] = A_c\sqrt{\frac{2}{N}} \cos\left(\frac{\pi(2m+1)}{2N}n\right),
\label{eq:dct_mod}
\end{equation}
for $n=0,\dots, N-1$. Notice that \eqref{eq:dct_mod} corresponds to \eqref{eq:prev_dct_mod} restricted to one tone and with no data regarding the DCT coefficients. The receiver can only demodulate the measurement $m$ and compute the desired function using the corresponding DCT approximation. Nevertheless, it is reasonable to assume that the receiver knows the function $f$, since it is generally the one requesting the communication. Notice that \eqref{eq:dct_mod} corresponds to the LoRa modulation (see \cite{Chiani2019}) using the DCT and not DFT basis.

After downconversion, filtering at $B_z^{max}=W(2N-1)/4$ and sampling at $f_s$, the signal at the input of the receiver is 
\begin{equation}
y[n] = hz[n] + w[n]\quad \text{for} \quad n=0,\dots, N-1,
\label{eq:signal_rx}
\end{equation}
where $w$ corresponds to complex AWGN samples. We consider perfect carrier and phase synchronization at downconversion so that $y[n]\in\mathbb{R}$. The demodulation process is the same as for \eqref{eq:prev_dct_mod}, this is, computing the DCT in the time domain, which allows to recover an estimate of the measurement $\hat{m}$. Since it corresponds to an $M$-ary Frequency Shift Keying (FSK) with $M=N$, the error in detecting $m$ can be approximated as
\begin{equation}
    P_e \approx (N-1)Q\left(\sqrt{\frac{A_c^2|h|^2}{N_o}}\right)
    \label{eq: error_prob}
\end{equation}

The performance in terms of MSE is intimately related to the error probability in \eqref{eq: error_prob}, and formulated as
\begin{equation}
    MSE(f,K) = \frac{1}{N}\sum_{k\not\in \mathcal{K}} F_k^2 + P_e\sum_{m=0}^{N-1} |f(m)-f(\hat{m})|^2
    \label{eq:mse_mean}
\end{equation}

The first term corresponds to the DCT approximation error, this is, the sum-power of the discarded coefficients. The second one corresponds to the MSE when the measurement is incorrectly estimated ($\hat{m}\neq m$), which is why it is weighted by the probability of miss-detection in $M$-ary FSK. This approximation is correct provided that $K$ is sufficiently large. Notice that the first term is controlled by the receiver at the desired degree of accuracy and vanishes very fast due to the energy compaction property of the DCT.

The communication properties of the LoRa modulation are highly convenient for the satellite context.
From an implementation perspective, \eqref{eq:dct_mod} reduces the complexity at the transmitter with respect to \eqref{eq:prev_dct_mod}, since the modulation only requires an FSK modulator and an ADC, which is already present in any sensor hardware. The transmission of a single frequency makes it a constant envelope modulation, which is bandwidth efficient. While the channel phase must be synchronized at the receiver, the transmitter does not need channel state information (CSIT) because the amplitude variations due to fading do not convey any loss of information. In this sense, it is also insensitive to amplitude variations regardless of the Doppler spread generated by high velocities of LEOs. In addition, the constant envelope is very welcome in satellite communications due to the use of high power amplifiers. Overall, its low complexity adapts very well to IoT and Sat-IoT frameworks, which incur very tight delay and energy constraints.

\subsection{Chirp Spread Spectrum}
LEO satellites are a promising approach for communicating with IoT devices in remote areas, as they allow to provide global coverage with reduced delays and losses. Nevertheless, ground to space communications exhibit severe channel degradation and IoT devices display tight energy constraints. In this respect, the LoRa modulation has gain relevance in its adaptability in IoT to satellite communications \cite{Wu2019}. It consists in an $M$-ary FSK coupled with a chirp spread spectrum (CSS) technique that allows to spread a narrowband signal over a wide bandwidth. This technique allows to develop long range communications while severely improving the receiver sensitivity by 20 dB. In the context of inter-satellite communications implementing the LoRa modulation allows to reduce the energy consumption and diminish the effect of interferences to other satellites. Besides, it has been shown that the LoRa modulation provides high immunity to the Doppler effect  generated by high speeds of LEO satellites \cite{Doroshkin2019}.

Consider the analytical signal of ${z[n]}$ in \eqref{eq:dct_mod}:
\begin{equation}
    z_{an}[n] = A_c\sqrt{\frac{2}{N}} \exp\left(j\frac{\pi(2m+1)}{2N}n\right)
\label{eq:dct_mod_analytical}
\end{equation}

The corresponding instantaneous frequency can be linearized by adding a term that increases with $n$:
\begin{equation}
    f_{LoRa}[n;m] = \frac{2m+1}{4N} + \frac{f_{mod}}{2N}n,
\label{eq:freq_inst_lora}
\end{equation}
where $f_{mod}$ allows to control the spreading factor of the chirped signal. To generate such a waveform, \eqref{eq:dct_mod_analytical} is multiplied by a chirp signal as
\begin{equation}
    z_{LoRa}[n] = z_{an}[n]\exp{\left(j\frac{\pi f_{mod}}{N}n^2\right)},
    \label{eq:lora_mod}
\end{equation}

Using this technique, the instantaneous frequency swipes over $f_{mod}/2$, which is not exact, but approximately the bandwidth occupied for $N\geq8$ \cite{Chiani2019}.

\section{LoRa for Type-based Multiple Access}
\begin{figure*}[t]
    \centering
    \includegraphics[width=0.9\textwidth]{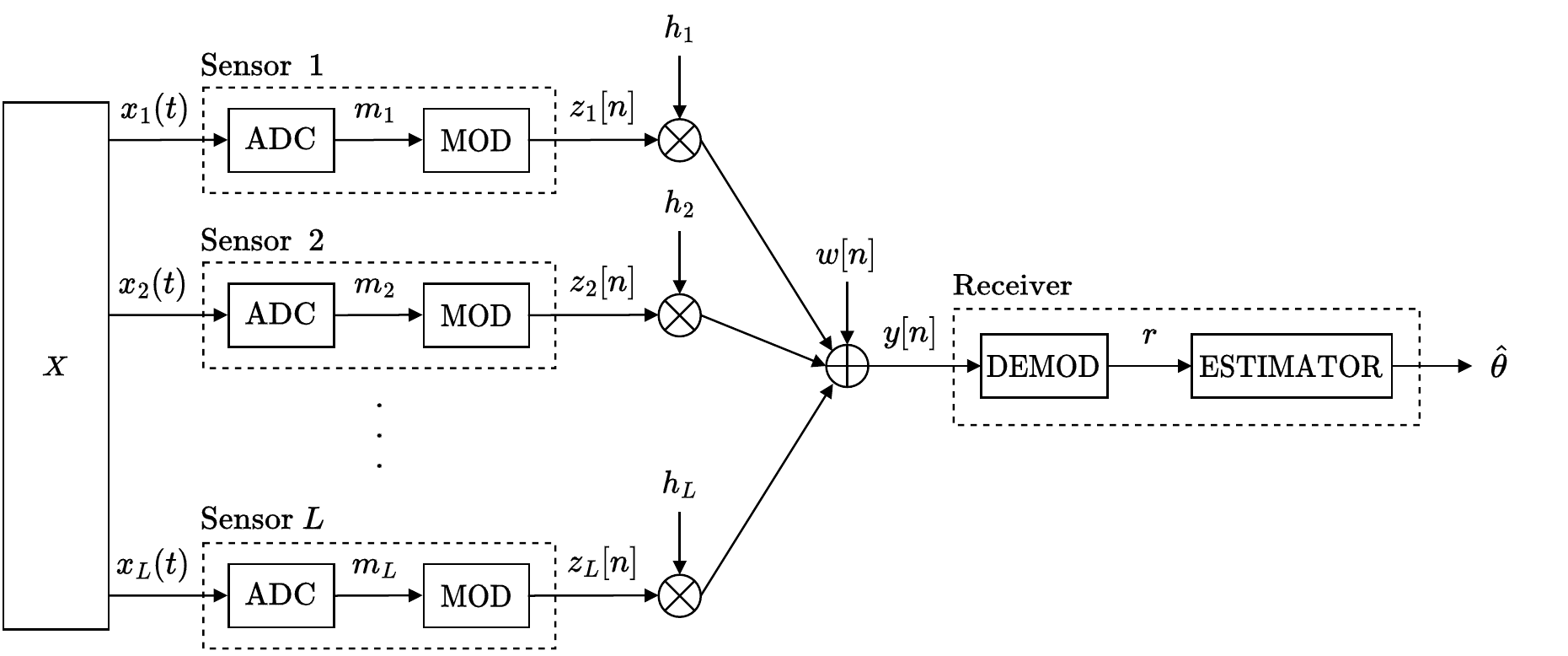}
  \caption{Multiple Access scheme for parameter estimation and function computation.}
  \label{fig:multiple_access}
\end{figure*}

Consider the scenario depicted in Fig. \ref{fig:multiple_access}, in which a set of $L$ sensors monitor a continuous random variable $X$ with a probability density function (pdf) in the domain $[0,N-1]\subset\mathbb{R}$. Each sensor $l$ gathers a realization and quantizes it, yielding the measurement $m_l$. Assuming a sufficiently large $N$, the probability mass function (pmf) generated over $[0,N-1]\subset\mathbb{N}$ corresponds to the pdf. The pmf is
$\textbf{p}_\theta\in\mathbb{R}^N$, where $\theta\in\Theta$ parameterizes the pdf and $\Theta\subset\mathbb{R}$ is the parameter space. The goal of the communication system is to provide a modulation that allows to obtain an estimate of the parameter $\hat{\theta}$ at the receiver.

One straightforward approach is to allocate orthogonal resources, which allows to recover all the noisy measurements at the receiver. Nevertheless, this procedure is not task-oriented and does not consider the semantic information in the data to optimize the resource utilization. For instance, using Time Division Multiple Access (TDMA) may be inadequate because the total transmission time scales with the number of sensors, but the LoS window for the satellite is limited.

An alternative for a massive access could be Non Orthogonal Multiple Access (NOMA), which has been considered in 5G and beyond. This access presents a high spectral efficiency if there is a proper power allocation and/or clustering at the transmitters, together with sequential interference cancelling at the receivers. However, as we are interested in computing a task rather than receiving information separatedly and reliably, we can avoid some of the requirements of NOMA, yet keeping the spectral efficiency gains due to the superimposed transmission of signals.

We propose using the LoRa modulation. The signal at the input of the receiver signal is
\begin{equation}
    y[n] = \sum_{l=1}^L A_c h_lz_l[n] + w[n],
    \label{eq:tbma}
\end{equation}
where
\begin{equation}
z_l[n] = A_c\sqrt{\frac{2}{N}} \cos\left(\frac{\pi(2m_l+1)}{2N}n\right)
\label{eq:dct_tbma}
\end{equation}
is the waveform transmitted by user $l$, namely expression \eqref{eq:dct_mod} with measurement $m_l$. For the sake of simplicity assume synchronization and a perfect channel ($h_l=1$). At the receiver side, after computing the DCT the signal and normalizing by $1/A_cL$, it results in:
\begin{equation}
    \textbf{r} = \frac{1}{L} [L_0,\dots,L_{N-1}]^T + [\tilde{w}_0,\dots,\tilde{w}_{N-1}]^T = \tilde{\textbf{p}}_\theta + \tilde{\textbf{w}}, 
    \label{eq:empirical_measure}
\end{equation}
where $\tilde{\textbf{p}}_\theta\in\mathbb{R}^N$ a vector whose $n$-th entry corresponds to the number of users $L_n$ that transmitted the measurement $n$ and $\tilde{w}_n$ is Gaussian noise with power $\sigma^2 = N_0/2A_c^2L^2$. Provided that $L>N$, the measurements $m_l$ are not necessarily different and $\textbf{r}$ corresponds to a nosy version of the empirical measure $\tilde{\textbf{p}}_\theta$.

In fact, the modulation presented in this work is a particular case of TBMA \cite{Mergen2006}, a semantic-aware multiple-access scheme in which orthogonal resources are only assigned to users whose data is informative; otherwise, when data is redundant, it falls under the same communication resource. In other words, resources are assigned to measurements, not users. In this respect, notice that TBMA relies on Over-the-air computing (AirComp) \cite{Nazer2007} as the concurrent transmission is exploited for all the transmitters that have the same measurement $m$. It harnesses interferences in a semantic approach in order to provide information at the receiver without the need of assigning orthogonal resources. AirComp provides a gain not perceived by orthogonal multiple access schemes that is proportional to the number of users, as the noise power in \eqref{eq:empirical_measure} decreases with $L^2$. Also, inherited from AirComp, TBMA is a bandwidth efficient scheme, since the number of orthogonal resources is upper bounded by $N$, while an orthogonal scheme scales linearly with the number of users.

In \cite{Mergen2006} it is shown that under perfect channel, $\textbf{r}$ converges in probability to $\textbf{p}_\theta$. This is, the empirical measure is a sufficient statistic of the random variable $X$ and the parameter $\theta$ can be estimated without loss of information. For large $L$, which can be assumed in an IoT scenario, the density of the first $N-1$ samples of $\textbf{r}$ is proportional to
\begin{equation}
    g(r(0),\dots,r(N-2))\propto \exp\left(-\frac{L}{2}\sum\limits_{n=0}^{N-1}\frac{(p_\theta(n)-r(n))^2}{p_\theta(n)}\right),
\end{equation}
where $r(n)$ and $p_\theta(n)$ are the $n$-th components of $\textbf{r}$ and $\textbf{p}_\theta$, respectively. The Maximum Likelihood estimator of a Gaussian distribution can be found in a closed-form expression, which is shown in \eqref{eq:ml_estimator}. Notice that this expression only holds for large $L$ and, thus, it corresponds to an asymptotic version of the ML estimator. 
\begin{equation}
    \hat{\theta} = \argmin_{\theta\in\Theta} \sum_{n=0}^{N-1}\frac{(p_\theta(n)-r(n))^2}{p_\theta(n)}
    \label{eq:ml_estimator}
\end{equation}


\subsection{Satellite Scenarios for TBMA}
In the following we motivate three different cases in which TBMA can be useful:
\begin{itemize}
    \item \textbf{Satellite Uplink}: 
    A set of ground IoT devices can communicate with a LEO when a ground infrastructure is not available. For a massive amount of devices, a classical orthogonal multiple access scheme is not suitable because the number of resources increases with the number of devices. For instance, FDMA incurs in large bandwidth consumption, while TDMA may be unattainable under the limited LoS window. There exist works in the literature that resort to NOMA altough they require complex receivers, which resort either to sequential interference canceller (SIC) or to spatial diversity multiple access (SDMA) \cite{Caus2021}, to separate the simultaneously received signals. In this way, the semantic nature of TBMA conveys bandwidth and energy efficiency.
    
    \item \textbf{Satellite Downlink}: For some constellation deployments several LEOs can perform LoS transmission simultaneously \cite{DELPORTILLO2019}. As in the uplink scenario, TBMA is advisable in order to reduce energy consumption and induce smaller delays.

    \item \textbf{Inter-satellite links}: In a Sat-IoT environment, LEOs can form a hierarchical computing structure \cite{Wei2019}. Similarly, when a swarm of satellites cannot communicate with a ground station due to lack of LoS, another satellite from a different orbit (e.g., GEO) can be used as a relay node. For either inter- and intra-orbit cases, inter-satellite link communications can incorporate TBMA to reduce the processing load towards the receiver.
\end{itemize}

\section{Numerical results}
\subsection{Experimental setup}
While it has been discussed its advantages in terms of delay and bandwidth requirements, in here we provide experimental evidence that LoRa-TBMA outperforms orthogonal schemes in task-related goals.
We explore the performance of TBMA in terms of $L$, as the number of users that the Sat-IoT scenarios can handle differs: while the uplink can support thousands of sensors, the downlink may handle up to 10 simultaneous LEOs \cite{DELPORTILLO2019}.

We consider the problem of estimating a distribution and its mean, specifically, a Gaussian random process. In TBMA the receiver knows the variance and uses the asymptotic ML estimator from \eqref{eq:ml_estimator}. Notice that the estimation may be conducted in parallel for all the parameter space, i.e. $\Theta=[0,N-1]$. We assume perfect time and phase synchronization.
As a benchmark we choose Time Division Multiple Access (TDMA), although any orthogonal scheme is identical in terms of performance. In TDMA, each transmitters sends its measurement using the LoRa modulation in a different time slot and the receiver computes the ML estimator (i.e., the empirical mean).


\subsection{Satellite Channel}
For all the three scenarios we assume that the LoS component is dominant, reinforced by the robustness of the chirp to multipath. We consider only the path loss, while the Doppler is compensated.
The signal to noise ratio (SNR) observed at the input of the receiver due to transmitter $l$ is
\begin{equation}
    \gamma_l = 
    \frac{P_{tx,l}\,h_l^2}{\sigma^2}= \frac{P_{tx,l}G_{tx}\,G_{rx}}{a_l\,\sigma^2},
\end{equation}
where $P_{tx,l}=A_{c,l}^2/N$ is the transmitted power of \eqref{eq:dct_mod}, $a_l$ is its corresponding path loss, $G_{tx}$ and $G_{rx}$ are the
transmitter and receiver antenna gains, and $\sigma_2$ is the noise power. The configuration for the scenarios is set according to \cite{Fernandez20}: The constellation is deployed at $600$ km, only satellites implement an array providing a gain of $15$ dB; the carrier frequency is set to $915$ MHz with a bandwidth of $500$ KHz; we assume $3$ dB for atmospheric, polarization and pointing losses, the latter in directional antennas. These configurations result in the SNR depicted in Fig. \ref{fig:sim_K100} 
for the uplink (UL), downlink (DL), LEO inter-satellite (ISL-LEO) and GEO inter-satellite (ISL-GEO).


\subsection{Performance analysis}

\begin{figure}[t]
\centering
\includegraphics[width=\columnwidth]{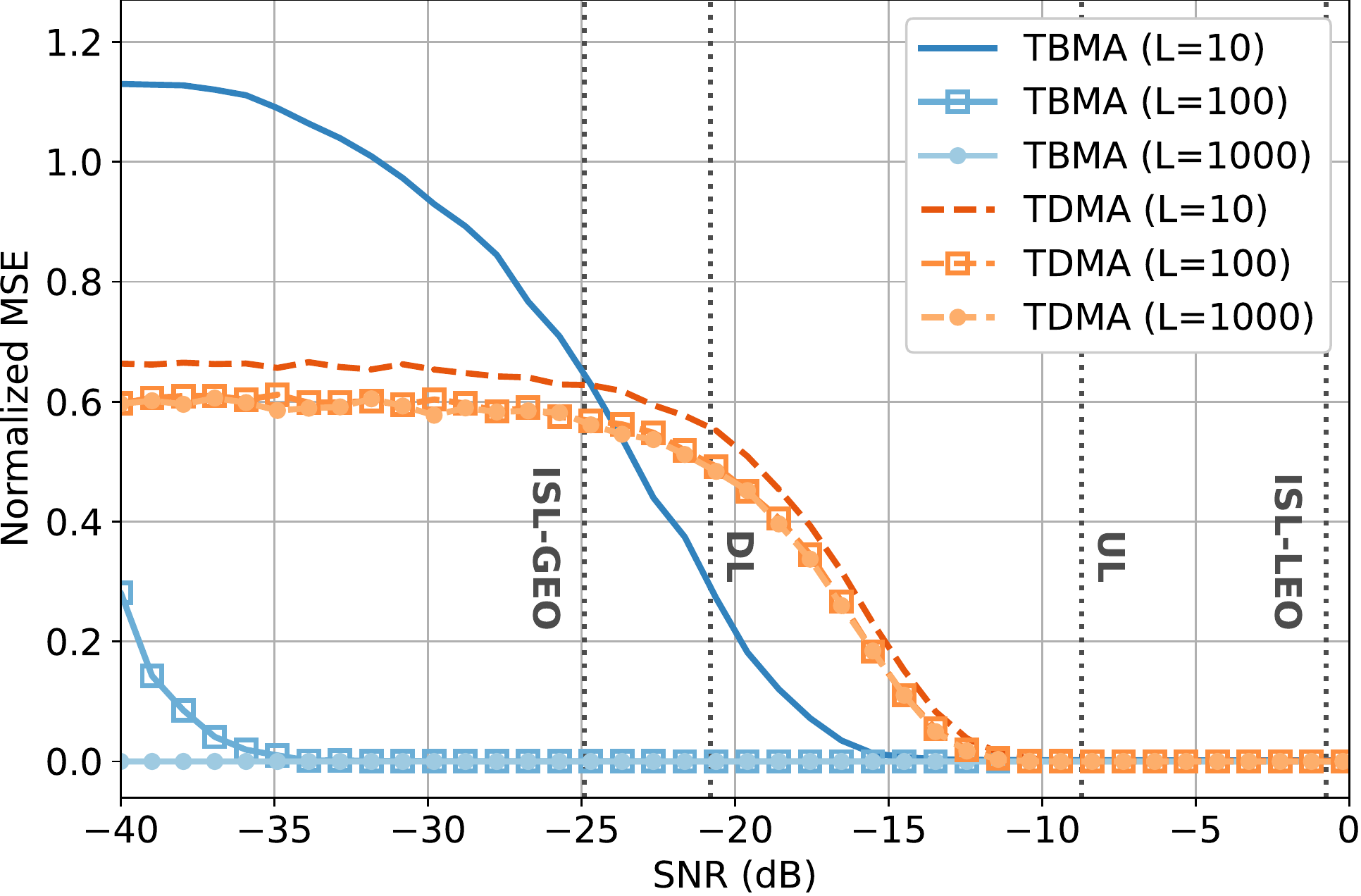}
\caption{Normalized MSE in estimating the mean of a random Gaussian process with TBMA and TDMA for $L=\{10,100,1000\}$ users for the different scenarios. In vertical dotted lines the working SNR for the different scenarios are indicated.}
\label{fig:sim_K100}
\vspace{-0 pt}
\end{figure}

\begin{figure}[t]
\centering
\includegraphics[width=\columnwidth]{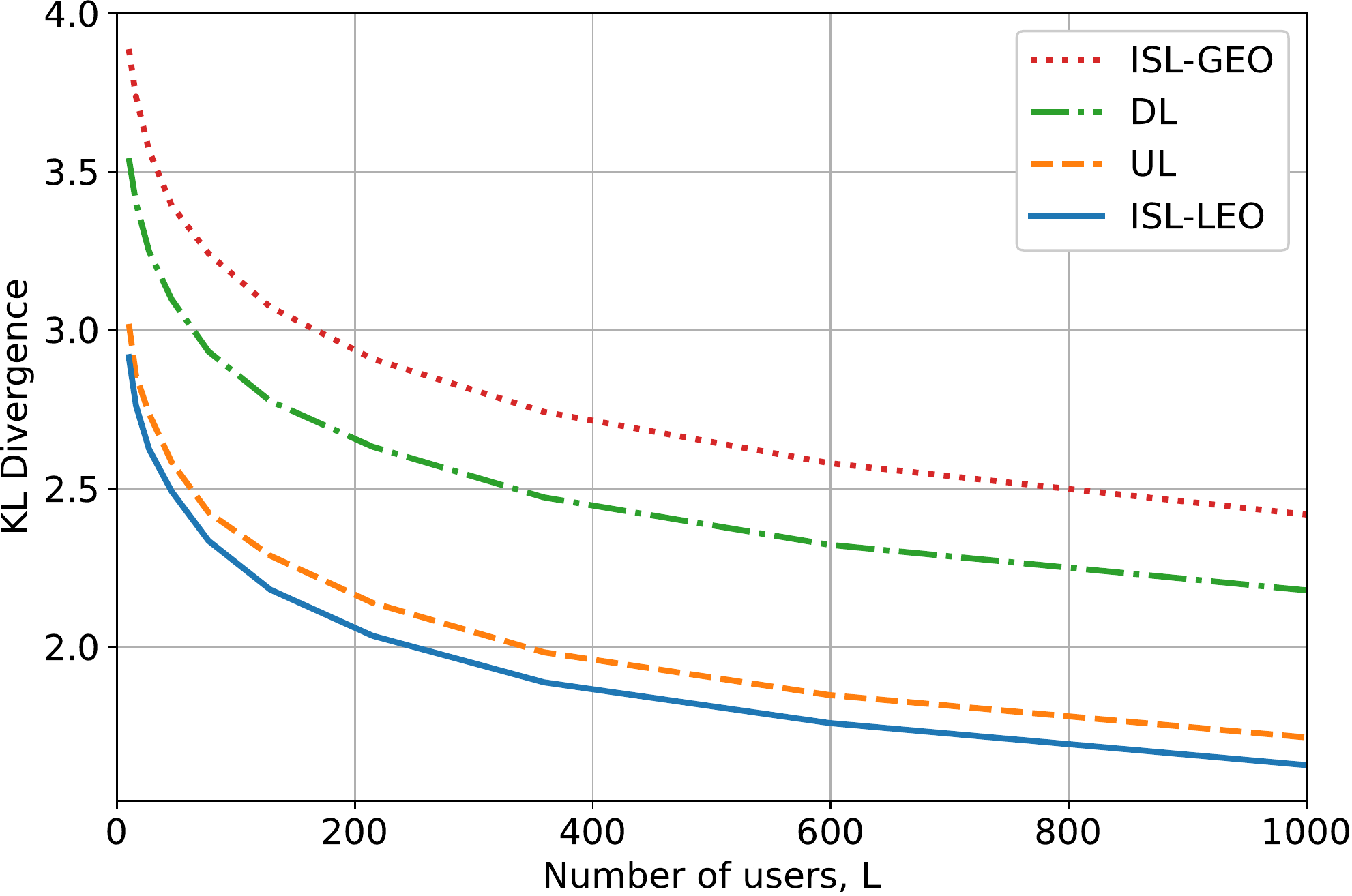}
\caption{KL divergence with respect to the true Gaussian Distribution with respect to the number of users in the 4 satellite scenarios.}
\label{fig:sim_KL}
\vspace{-0 pt}
\end{figure}

Fig. \ref{fig:sim_K100} shows the normalized MSE of TBMA and TDMA for $L=\{10,100, 1000\}$ transmitters and $N=256$.
As expected, TBMA exhibits a gain that increases with the number of transmitters $L$, while TDMA does not.
While TBMA always provide a gain in delay due to AirComp (i.e., reducing the delay by a factor of $L$), it also enables to reduce the transmitted power for a large $L$. This allows to remain above the sensitivity of the LoRa receiver and work at a smaller SNR than the one depicted in Fig. \ref{fig:sim_K100}. All that without compromising the MSE. Fig. \ref{fig:sim_KL} shows the KL divergence with respect to the true distribution for an increasing number of users. When more users are available, the received distribution approaches the true and better estimations can be computed from the data.


\section{Conclusions}
In this paper we propose to use the LoRa modulation for the purpose of estimation in Sat-IoT. It shows that angular modulations, which are very useful in satellite communications, can also benefit from AirComp. Specifically, we proposed to develop a LoRa-TBMA scheme, which allows to estimate a parameter of distributed measurements by exploiting the semantics of the data. The experimental results demonstrate the gains of LoRa-TBMA in terms of communication resources (i.e., delay, bandwidth and power), while outperforming orthogonal schemes in  task-oriented communications.


\bibliographystyle{IEEEbib}
\bibliography{refs}

\begin{thebibliography}{10}

\bibitem{Qu2017}
Zhicheng Qu, Gengxin Zhang, Haotong Cao, and Jidong Xie,
\newblock ``{LEO} satellite constellation for {Internet of Things},''
\newblock {\em IEEE Access}, vol. 5, pp. 18391--18401, 2017.

\bibitem{Akyildiz2019}
Ian~F. Akyildiz and Ahan Kak,
\newblock ``The internet of space things/cubesats: A ubiquitous cyber-physical
  system for the connected world,''
\newblock {\em Computer Networks}, vol. 150, pp. 134--149, 2019.

\bibitem{Nazer2007}
Bobak Nazer and Michael Gastpar,
\newblock ``Computation over multiple-access channels,''
\newblock {\em IEEE Transactions on Information Theory}, vol. 53, no. 10, pp.
  3498--3516, 2007.

\bibitem{Qin2021}
Zhijin Qin, Xiaoming Tao, Jianhua Lu, and Geoffrey~Ye Li,
\newblock ``Semantic communications: Principles and challenges,''
\newblock {\em arXiv preprint arXiv:2201.01389}, 2021.

\bibitem{Chiani2019}
Marco Chiani and Ahmed Elzanaty,
\newblock ``On the {LoRa} modulation for {IoT}: Waveform properties and
  spectral analysis,''
\newblock {\em IEEE Internet of Things Journal}, vol. 6, no. 5, pp. 8463--8470,
  2019.

\bibitem{Gost23}
Marc~M. Gost, Ana Pérez-Neira, and Miguel~Ángel Lagunas,
\newblock ``{DCT-based air interface design for function computation},''
\newblock {\em IEEE Open Journal of Signal Processing}, pp. 1--9, 2023.

\bibitem{Mergen2006}
G.~Mergen and L.~Tong,
\newblock ``Type based estimation over multiaccess channels,''
\newblock {\em IEEE Transactions on Signal Processing}, vol. 54, no. 2, pp.
  613--626, 2006.

\bibitem{Wu2019}
Tingwei Wu, Dexin Qu, and Gengxin Zhang,
\newblock ``Research on {LoRa} adaptability in the {LEO} satellites internet of
  things,''
\newblock in {\em 2019 15th International Wireless Communications \& Mobile
  Computing Conference (IWCMC)}, 2019, pp. 131--135.

\bibitem{Doroshkin2019}
Alexander~A. Doroshkin, Alexander~M. Zadorozhny, Oleg~N. Kus, Vitaliy~Yu.
  Prokopyev, and Yuri~M. Prokopyev,
\newblock ``Experimental study of {LoRa} modulation immunity to doppler effect
  in {CubeSat} radio communications,''
\newblock {\em IEEE Access}, vol. 7, pp. 75721--75731, 2019.

\bibitem{Caus2021}
Marius Caus, Ana Perez-Neira, and Eduard Mendez,
\newblock ``Smart beamforming for direct {LEO} satellite access of future
  {IoT},''
\newblock {\em Sensors}, vol. 21, no. 14, 2021.

\bibitem{DELPORTILLO2019}
Inigo {del Portillo}, Bruce~G. Cameron, and Edward~F. Crawley,
\newblock ``A technical comparison of three low earth orbit satellite
  constellation systems to provide global broadband,''
\newblock {\em Acta Astronautica}, vol. 159, pp. 123--135, 2019.

\bibitem{Wei2019}
Junyong Wei, Jiarong Han, and Suzhi Cao,
\newblock ``Satellite iot edge intelligent computing: A research on
  architecture,''
\newblock {\em Electronics}, vol. 8, no. 11, 2019.

\bibitem{Fernandez20}
Lara Fernandez, Joan~Adria Ruiz-De-Azua, Anna Calveras, and Adriano Camps,
\newblock ``Assessing {LoRa} for satellite-to-{Earth} communications
  considering the impact of ionospheric scintillation,''
\newblock {\em IEEE Access}, vol. 8, pp. 165570--165582, 2020.

\end{thebibliography}

\end{document}